\documentclass[aps,preprint,showpacs,showkeys]{revtex4}
\usepackage{graphicx}
\usepackage{amsmath}
\usepackage{amsfonts}
\usepackage{amsthm}
\usepackage{mathrsfs}
\usepackage{amssymb}
\usepackage{graphicx}
\usepackage{epsfig}
\usepackage{hyperref}
\usepackage{amscd}
\begin{document}
\newcommand{\be}{\begin{equation}}
\newcommand{\ee}{\end{equation}}
\newcommand{\bea}{\begin{eqnarray}}
\newcommand{\eea}{\end{eqnarray}}

\title{Gravitational field  of  slightly deformed naked singularities%\thanksref{t1}
}
\author{Saken Toktarbay$^{1,2,5}$ , Hernando Quevedo$^{1,3,4}$, Medeu Abishev$^{1,2}$
and Aray Muratkhan$^{1,2}$ %etc.
}

\affiliation
{ $^1$ Institute for Experimental and Theoretical Physics, \\ Al-Farabi Kazakh National University, Almaty, 050040, Kazakhstan \\ 
$^2$ Institute of Nuclear Physics, 1 Ibragimova St.
Almaty, 050032, Kazakhstan \\
$^3$Instituto de Ciencias Nucleares, Universidad Nacional Aut\'onoma de M\'exico, AP 70543, M\'exico, DF 04510, Mexico \\
$^4$ Dipartimento di Fisica and ICRA, Universit\`a di Roma ``La Sapienza", I-00185 Roma, Italy \\
$^5$ Department of Physics, Kazakh National Women's Teacher Training University, Almaty 050000, Kazakhstan 
}

\date{}

\begin{abstract}
We derive a particular approximate solution of Einstein equations, describing the gravitational field of a mass distribution that  slightly deviates from spherical symmetry. The deviation is described by means of a quadrupole parameter that is responsible for the appearance of a curvature singularity, which is not covered by a horizon. We investigate the motion of test particles in the gravitational field of this naked singularity and show that the quadrupole parameter affects the properties of Schwarzschild trajectories. By investigating radial geodesics, we find that no effects of repulsive gravity are present. We interpreted this result as indicating that repulsive gravity is non-linear effect. \end{abstract}

\keywords{compact object, quadrupole moment, approximate solution}
\maketitle

\section{\label{sec:intro}Introduction}

Although the existence of naked singularities in Nature is the subject of intense debate nowadays, it has been well established that Einstein field equations for the gravitational field allow solutions that can be interpreted as describing naked singularities. In particular, black hole solutions are characterized by the existence of naked singularity counterparts \cite{kerr}. However, it seems that the particular choice of the physical parameters, which is necessary for the formation of naked singularities, is difficult to be realizable in Nature. Indeed, a rotating naked singularity needs a specific angular momentum that must be greater that its mass (in geometric units), a condition that  probably cannot be fulfilled in realistic configurations because it would imply such a high angular velocity that the object would destroy itself before reaching it \cite{mtw}.

There is, however, a simpler way to generate naked singularities, namely, by considering mass distributions with quadrupole moment \cite{quev11}. Indeed, from the point of view of multipole moments, the uniqueness theorems prove that black holes can have only mass monopole and angular momentum dipole \cite{heusler}. Consequently, the addition of a quadrupole to a mass distribution, even in the static case, would imply that the corresponding gravitational field describes a naked singularity. Consequently, a simple shape deviation from spherical symmetry in a mass distribution leads to the appearance of naked singularities. 
In previous works \cite{quev11,Quevedo2016,Boshkayev2016,Boshkayevpxa}, we used  a particular static quadrupolar solution \cite{zipoy,voorhees,malafarina05,solutions}
to study the physical properties of naked singularities.

There are several  solutions of Einstein field equations that can be used to describe the exterior gravitational field of a static mass distribution with quadrupole moment \cite{aqs18}. In the limiting case of vanishing quadrupole, they reduce to the spacetime of a Schwarzschild black hole.
An interesting characteristic of all of them is that the outermost naked singularity is a sphere with radius $r=2m$, where $m$ is the mass of the gravitational source. This could be interpreted intuitively as if the presence of the quadrupole causes the destruction of the regular horizon turning it into a singular hypersurface. 

In this work, we show that this is not always the case. Indeed, we will derive  a new solution, whose singularity is located on a sphere of radius $r=m$. This means that in this case the quadrupole completely destroys the regular horizon at $r=2m$, but generates a new special hypersurface at $r=m$ which contains a singularity. To derive this new solution, we use the fact in realistic situations we expect that compact objects deviates only slightly from spherical symmetry. This implies that the quadrupole can be considered as a small quantity. With this in mind, we investigate a particular approximate line element, which is valid only up to the first order in the quadrupole parameter. Then, we find the general solution of the corresponding Einstein vacuum field equations and show that a particular solution is characterized by a curvature singularity located on a sphere of radius $r=m$.

  This work is organized as follows. In Sec. \ref{sec:exq}, we consider a line element that is especially adapted to the study of interior and exterior solutions. We derive the general field equations for the case of vacuum gravitational fields. In Sec. \ref{sec:gvs}, we find the most general solution that is linear in the quadrupole moment. We select a particular case that is characterized by the presence of naked singularity at a distance $r=m$ from the origin of coordinates.
   We also calculate the Newtonian limit of the new approximate solution and show that it corresponds to a mass distribution with a small quadrupole. In Sec. \ref{sec:geo}, we investigate the motion of test particles in the spacetime described by the approximate solution. In general, we find that the quadrupole affects the behavior of Schwarzschild orbits.  By analyzing the behavior of free falling particles we show that no effects associated with the presence of repulsive gravity can be detected in contrast to repulsive effects found previously in the case of exact solutions with quadrupole. We conclude that repulsive gravity in naked singularities is a non-linear phenomenon. 
     Finally, Sec. \ref{sec:con} contains a summary of our results.

%%%%%%%%%%%%%%%%%%%%%%%%%%%%%%%%%%
%%%%%%%%%%%%%%%%%%%%%%%%%%%%%%%%%%

%%%%%%%%%%%%%%%%%%%%%%%%%%%%
%%%%%%%%%%%%%%%%%%%%%%%%%%%
 
\section{\label{sec:exq} Line element and field equations}

The search for and investigation of physically meaningful solutions of Einstein equations begins with the choice of an appropriate line element. In the case of static gravitational fields of deformed mass distributions, one can assume that the fields preserve axial symmetry. Moreover, if we are interested in describing the gravitational field outside as well as inside the mass distribution, it is convenient to choose a line element that can be used in both cases. 
In a previous work \cite{approxi_2021}, we found interior perfect-fluid solutions that can be matched with exterior vacuum solutions under the assumption that the quadrupole moment is small. From a physical point of view, this implies that the mass distribution is only slightly deformed. We were able to find a particular line element that can be used to search for interior and exterior approximate solutions. It can be written as 
\begin{eqnarray}
ds^2 =&&  e^{2 \nu} (1+qa) dt^2 - (1+qc+qb)\frac{dr^2}{1-\frac{2\tilde m}{r}} \nonumber \\ &&-
(1+qa+qb)r^2d\theta^2
-(1-qa)r^2\sin^2\theta d\varphi^2\ ,
\label{apin1}
\end{eqnarray} 
where the set of $(t,r,\theta,\varphi)$ can be interpreted as polar coordinates in the limiting case $q\rightarrow 0$. Moreover,  the functions $\nu=\nu(r)$, $a=a(r)$, $c=c(r)$, $\tilde m=\tilde m(r)$, and $b=b(r,\theta)$ are arbitrary. 

In the particular case $q=0$, the above line element can be used to describe the exterior Schwarzschild solution 
\be 
\tilde m = m =const,\ e^{2\nu} = 1-\frac{2m}{r}\ ,
\ee
and the interior perfect fluid Schwarzschild metric with 
\be 
\tilde m = 4 \pi \rho R^{3} \ , \ 
p       =   \frac{3m}{4\pi R^3} \frac{\left[ f(r) - f(R) \right ] }
{\left[    3  f(R) - f(r) \right ]}\ ,
\ee
with
\be  e^{2\nu} = \bigg[ \frac{3}{2} f\left(R \right) -\frac{1}{2} f\left(r \right) \bigg]^2,\ \ 
              f\left(r \right)= \sqrt{1-\frac{2mr^2}{R^3}} \ee
where $\rho = const$ and $p=p(r)$ are the density and pressure of the fluid, respectively.

Furthermore, for $q\neq 0$ the line element (\ref{apin1}) contains the approximate (up to the first order in $q$) quadrupolar metric ($q-$metric), which has been interpreted as the simplest generalization of the Schwarzschild metric that includes a quadrupole moment \cite{quev11}. In this limit,  $q$ has been interpreted as the quadrupole parameter. 

Here, we will use the advantages of the line element (\ref{apin1}) to search for more general approximate solutions. Then, the Einstein vacuum field equations, up to the first order in $q$, can be written as
\begin{equation}
\tilde m_{,r} = 0 \quad {\rm i.e.} \quad \tilde m = m = const.\ ,
\end{equation}
\begin{equation}
\nu_{,r}=\frac{ {m}}{r \left(r-2 {m} \right)}\ ,
\label{nuv}
\end{equation}
\begin{equation}
(r-m)(a_{,r}-c_{,r}) + (a-c) = 0\ ,
\label{acv}
\end{equation}
\begin{eqnarray} \nonumber
2 &&r \left(r-2m \right)a_{,rr} + \left(3r-m \right) a_{,r} + \left(r-3m \right) c_{,r} \\ &&- 2 \left(a-c \right)=0,
\end{eqnarray} 
\begin{eqnarray} \nonumber
r &&\left(r-2m \right) b_{,rr}+ b_{,\theta \theta} + \left(r-m \right) b_{,r} -2 \left(r-2m \right) c_{,r} \\ &&+ 2 \left(a-c \right)=0 ,
\end{eqnarray} 
\begin{eqnarray} \nonumber 
&& \left( r^2 -2 mr + m^2 \sin^2 \theta  \right) b_{,\theta} + 2 r \left(r-2m \right) \nonumber \\ && \times \left( m a_{,r}-  a+c \right) \sin \theta \cos \theta =0 ,
\end{eqnarray}
\begin{eqnarray} \nonumber 
&&\left( r^2 -2 m r + m^2 \sin^2 \theta  \right)   b_{,r}+ 2 \left(r-2 m \right) \nonumber \\ && \times \left(r- m \sin^2 \theta \right) a_{,r}+2 \left(r-m \right) \left(a-c \right) \sin^2 \theta =0 ,
\label{eqb}
\end{eqnarray}
where a comma represents partial differentiation with respect to the corresponding coordinate. For simplicity, we replaced the solution of the first equation $\tilde m = m = const.$ in the remaining equations.

%%%%%%%%%%%%%%%%%%%%%%%%%%%%%%%%%%%%%%%%%%%%
%%%%%%%%%%%%%%%%%%%%%%%%%%%%%%%%%%%%%%%%%%%%%
\section{\label{sec:gvs} Approximate solutions}

We now investigate the system of partial differential equation (\ref{nuv})-(\ref{eqb}). 
Equations (\ref{nuv}) and (\ref{acv}) can be integrated and yield
\begin{equation}
\nu = \frac{1}{2} \ln\left(1-\frac{2m}{r}\right) + \alpha_1 \ ,\quad  a-c = \frac{\alpha_2 m^2}{(r-m)^2} \ ,
\label{sol_nuac}
\end{equation}
where $\alpha_1$ and $\alpha_2$ are dimensionless integration constants. It turns out that
the remaining system of partial differential equations can be integrated in general and yields
\begin{equation}
a= - \frac{\alpha_2 m}{r-m} + \frac{1}{2}\left(\alpha_3 - {\alpha_2}\right) \ln\left(1-\frac{2m}{r}\right) + \alpha_4\ ,
\label{vsola}
\end{equation}
\begin{equation}
c= - \frac{\alpha_2 m r}{(r-m)^2} + \frac{1}{2}\left(\alpha_3 - {\alpha_2}\right) \ln\left(1-\frac{2m}{r}\right) + \alpha_4\ ,
\label{vsolc}
\end{equation}
\begin{eqnarray} \nonumber
&b=& \frac{2\alpha_2 m}{r-m} -  \left(\alpha_3 - {\alpha_2}\right) 
\bigg[ \ln 2 \\ &+& \ln\left(1-\frac{2m}{r} + \frac{m^2\sin^2\theta}{r^2} \right)\bigg] + \alpha_5 \ ,
\label{vsolb}
\end{eqnarray}
where $\alpha_3$, $\alpha_4$, and $\alpha_5$ are dimensionless integration constants. We can see that the general approximate exterior solution with quadrupole moment is represented by the 5-parameter family of solutions (\ref{sol_nuac})--(\ref{vsolb}).
In this general solution, the additive constants $\alpha_4$ and $\alpha_5$ can be chosen such that at infinity the solution describes the Minkowski spacetime in spherical coordinates. This means that non asymptotically flat solutions are also contained in the above general solution. 

\begin{figure}%
\includegraphics[scale=0.4]{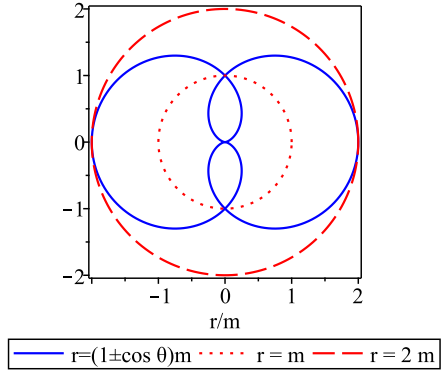}%
\caption{Curvature singularities of the general metric.
}% 
\label{fig1}%
\end{figure}

To partially investigate the physical meaning of this solution, we calculate the Kretschmann scalar $K= R_{abcd} R^{abcd}$. We obtain
\begin{eqnarray} \nonumber
&K=& \frac{48 m^2}{r^6} \Bigg \{1+ q \Bigg[  
\left( \alpha_2 -\alpha_3 \right) \Big[ \ln \left( 1- \frac{2m}{r}\right) \nonumber \\ 
&-& 2 \ln \left( 1- \frac{2m}{r} + \frac{m^2}{r^2} \sin^2{\theta} \right) - 2\ln2 \Big] \nonumber \\ 
&+\alpha_{2}& A_{1}+ \alpha_{3} A_{2} -2 \left( \alpha_4 + \alpha_5 \right) +\alpha_{2} A_{3} \Bigg] +\mathcal{O}(q^2)\Bigg\}
\end{eqnarray}
where 
 \begin{eqnarray} \nonumber 
     A_{1}&=&-\frac{1}{6 \left( m^2 \sin^2 \theta-2mr +r^2 \right) \left(m-r \right)^3 m^2} \nonumber \\ &\times& \bigg[ \left(12m^3-11m^2r-14m r^2+11r^3 \right)m^4 \cos^2 \theta  \nonumber \\  &-&\left(12m^3-17m^2r-2m r^2+5 r^3\right) m^2 \left(m-r\right)^2 \bigg]
 \end{eqnarray}
 
  \begin{eqnarray}
     A_{2}=  \frac{2 m^2 \sin^2 \theta - 3 m r+r^2}{ m^2 \sin^2 \theta-2mr +r^2}
 \end{eqnarray}
 
\begin{eqnarray}
A_{3}= - \frac{1}{6} \frac{r \big[ 7 \left(m-r \right)^2-2 r^2\big]}{ \left(m-r \right)^3}
 \end{eqnarray}
 where the term proportional to $q^2$ has been neglected. We can see that this approximate spacetime is characterized by the presence of three different curvature singularities located at 
 \be 
 r=0\ ,\quad r=m\,  \quad r=2m\ , \quad r=m(1\pm \cos\theta)\ .
 \ee
 The geometric structure of the curvature singularities of the solutions (\ref{sol_nuac}), (\ref{vsola}), (\ref{vsolc}) and (\ref{vsolb}) is illustrated in Fig.\ref{fig1}.

Another interesting particular case corresponds to the choice
\begin{equation} 
\alpha_1 =0 \ ,\quad \alpha_3 = {\alpha_2}\ ,\quad \alpha_4= 0\ ,\quad \alpha_5 = 0\ ,
\end{equation}
which leads to the line element
\begin{eqnarray} \nonumber
ds^2 &= & \left(1-\frac{2m}{r}\right)\left(1-\frac{q\alpha_2 m}{r-m}\right) dt^2 \nonumber \\  & & -\left[1+  \frac{q\alpha_2 m (r-2m)}{(r-m)^2} \right] \frac{dr^2}{1-\frac{2m}{r}} \nonumber \\ & & - \left(1 + \frac{q\alpha_2 m}{r-m} \right) r^2(d\theta^2+\sin^2\theta d\varphi^2) 
\label{appsol1}
\ .
\end{eqnarray}
This is an asymptotically flat approximate solution with parameters $m$, $q$ and $\alpha_2$. Since $\alpha_2$ appears always in combination with $q$, it can be absorbed in the definition of $q$. We, therefore, set $\alpha_2=1$ without loss of generality. 
The singularity structure can be found by analyzing the Kretschmann invariant,  which in this case reduces to
\begin{equation}
K= \frac{48 m^2}{r^6}\left(1+ q  \frac{r-4m}{r-m} + \mathcal{O}(q^2)\right)\ .
\label{Kretsch_q}
\end{equation}
We see that there are only two singularities, which are located  at $r=0$ and  $r=m$. This is an interesting property of  this solution because all the remaining solutions contained in (\ref{sol_nuac})--(\ref{vsolb}) are singular on the hypersurface $r=2m$. Moreover, all the known exact solutions with quadrupole turn out to be singular at $r=2m$ \cite{aqs18}. To our knowledge, the solution (\ref{appsol1}) is the only one in which the outermost singularity is located inside the sphere with $r=2m$. This means that the spacetime is well defined behind in the interval $r\in (m,2m]$. We are interested in investigating the properties of this spacetime near the singularity $r=m$.

To further analyze the physical meaning of the solution (\ref{appsol1}), we calculate the corresponding Newtonian limit. To this end, we perform a coordinate transformation of the form $(r,\theta)\rightarrow (R,\vartheta)$ defined by the equations \cite{approxi_2021,Mashhoon2018}
\begin{eqnarray}  \nonumber
r&=& R \bigg[1- q \frac{m}{R} \big[ 1+ \frac{m}{R} \left( \beta_{1} + \sin^2 \vartheta \right)  \\ &+& \frac{m^2}{R ^2} \left( \beta_{2} - \sin^2 \vartheta \right)+ .. \big] \sin^2 \vartheta \bigg]\label{trans_r} ,
\end{eqnarray}
and
\begin{equation}
\theta = \vartheta -q \frac{m^2}{R^2} \left(1+ 2 \frac{m}{R} + . . . \right) \sin \vartheta \cos \vartheta
\label{trans_theta},
\end{equation}
where the $\beta_1$ and $\beta_2$ are  constants and we have neglected terms of the order higher that $m^3 / R^3 $. Inserting the above coordinates into the metric (\ref{appsol1}), we obtain the approximate line element
\bea  \nonumber 
ds^2&=& \left( 1+ {2 \Phi} \right)dt^2 - \frac{dR^2 }{1+ {2 \Phi}} \\ & -& U \left( R, \vartheta \right) R^2 \left( d\vartheta^2 + \sin^2 \vartheta d \varphi^2 \right), 
\label{Newline}
\eea 
with  
\begin{equation}
    \Phi= -\frac{GM}{R}+ \frac{GQ}{R^3} P_{2}\left( \cos \vartheta \right),
\end{equation}
\begin{equation}
    U \left( R, \vartheta \right)= 1- 2 \frac{GM}{R^3} P_2(\cos \vartheta),
\end{equation}
where $P_2(\cos \vartheta)$ is the Legendre polynomial of degree 2, and we have chosen the free constants as $\alpha_{2}=2$, $\beta{_1}=1/3 $, and $\beta{_2}=5/3 $.

We recognize the metric (\ref{Newline}) as the Newtonian limit of general relativity, where $\Phi$ represents the Newtonian potential. Moreover, the constants  
\begin{equation}
    M= \left( 1+q \right)m, \ \ \ Q=\frac{2}{3} q m^3,
\end{equation}
can be interpreted as the Newtonian mass and quadrupole moment of the corresponding mass distribution. 

%Notice that in order to obtain the Newtonian limit (\ref{Newline}), it is necessary to apply the coordinate transformation (\ref{trans_r}), which relates $r$ with $R$ and $\vartheta$. This shows that $r$ cannot interpreted as a radial coordinate and surfaces with $r=const$ do not correspond to spheres.

%%%%%%%%%%%%%%%%%%%%%%%%%%%%%%%%
%%%%%%%%%%%%%%%%%%%%%%%%%%%%%%%%%
\section{\label{sec:geo} Motion of test particles}

Consider the trajectory $x^{a}(\tau)$ of a test particle with 4-velocity   $u^{a} =dx^{a}/d\tau = {\dot x}^{a}$.  
The  4-moment  $p^a = \mu {\dot x}^{a} $ of  the particle can be  normalized so  that the equations and constraint for geodesics are given as

\begin{equation}
    \ddot{x}^{a}+ {\Gamma^{a}}_{bc} \dot{x}^{b} \dot{x}^{c}=0
    \label{eq_geodesics}
\end{equation}

   \begin{equation}
    g_{a b} \dot x^{a} \dot x^{b} = \epsilon,\label{eq1:normomentum} 
   \end{equation}
   where $\epsilon =0, 1, -1$ for null,  timelike, and  spacelike curves, respectively \cite{Pugliese:2010ps}.  
For  the  approximate metric (\ref{appsol1}) we  obtain from (\ref{eq_geodesics}) that geodesics are determined, in general, by the following set of equations 
\begin{eqnarray}
\ddot{t}+ m \bigg[ 
\frac{2}{r \left(r-2m \right)} + \frac{q}{\left( r-m\right)^2} \bigg]\dot{t} \dot{r}=0 \ ,
\label{eq_time}
\end{eqnarray}

     \begin{eqnarray} \nonumber
    && \ddot{r}+ \bigg[ 
    1 + \frac{q\left( 6m^2-6mr+r^2\right)}{2\left( r-m\right)^2}
    \bigg]\frac{m \left(r-2m \right)\dot{t}^2}{r^3}
        \nonumber \\
     &&- \frac{m}{2} \bigg[ \frac{2}{r \left(r-2m \right)}+ \frac{q \left( r-3m\right)}{\left( r-m\right)^3}
     \bigg] \dot{r}^2 \nonumber \\
     &&- \bigg[ r-2m - \frac{qm \left(r-2m \right)^2}{ \left(r-m \right)^2} \bigg] 
     \dot{\theta}^2 \nonumber \\
     && - \left(r-2m \right) \sin^2 \theta \bigg[ 
     1- \frac{qm \left(r-2m \right)}{\left( r-m\right)^2}
     \bigg]
     \dot{\varphi}^2 =0 \ ,
     \end{eqnarray}
\begin{equation}
    \ddot \theta -\sin \theta \cos \theta \dot{\varphi}^2+ \left( \frac{2}{r}-\frac{qm}{ \left( r-m\right)^2}\right) \dot{r} \dot{\theta} =0 .
    \label{eq_theta}
\end{equation}

\begin{eqnarray}
\ddot {\varphi} + \Bigg[
\left( \frac{2}{r} - \frac{qm}{\left( r-m\right)^2} \right) \dot{r} + \frac{2 \cos \theta}{\sin \theta} \dot{\theta}
\bigg]\dot{ \varphi}=0 \ ,
\label{eq_phi}
\end{eqnarray}

The  4-moment  $p^a = \mu {\dot x}^{a} $ of  the particle can be  normalized so  that 
   \begin{eqnarray}
    g_{a b} \dot x^{a} \dot x^{b} = \epsilon,\label{eq:normomentum} 
   \end{eqnarray}
where $\epsilon =0, 1, -1$ for null,  timelike, and  spacelike curves, respectively \cite{pqr11a}. Then, for  the  approximate metric (\ref{appsol1}) we  obtain from (\ref{eq:normomentum}) that 
\begin{eqnarray}
  \left(1- \frac{q m^2}{\left(r-m \right)} \right){\dot r}^2 = {\tilde E}^2 - \Phi^2,\label{eq:genequat}
   \end{eqnarray}
where
\bea \nonumber
V_{eff}& =& \Phi^2 =   \left(1- \frac{2m}{r} \right) \bigg[ 
r^2 \dot{\theta}^2 + \left( \epsilon + \frac{2l^2}{r^2 \sin^2 \theta } \right) \\ &\times& \left(1- \frac{qm}{r-m} \right)- \frac{l^2}{r^2 \sin^2 \theta}
\bigg] 
  \label{veff}
   \eea 
   is the effective potential 
and  we  have used the expression for  the energy $E=\mu \tilde E$ and the angular  moment $l=\mu \tilde l$ of  the test particle which are constants of  motion 
    \begin{equation}
    E = g_{a b}\xi^{a}_{t}p^{b}=  \left(1-\frac{2m}{r}\right)\left(1-\frac{q m}{r-m}\right) \mu {\dot t},
    \label{cons_m_E}
     \end{equation}
     
     \begin{equation}
    l = - g_{a b}\xi^{a}_{\varphi}p^{b} = \left(1+  \frac{q m}{r-m}\right) r^2 \sin^2\theta \mu {\dot \varphi},
    \label{cons_m_l}
     \end{equation}
associated with the  Killing vector  fields $\xi_t= \partial_t$ and $\xi_{\varphi}= \partial_{\varphi}$, respectively. For the sake of simplicity we set $\mu=1$ so that $\tilde E = E$ and $\tilde l = l$.

\begin{figure}%
\includegraphics[width=7cm, height=5cm]{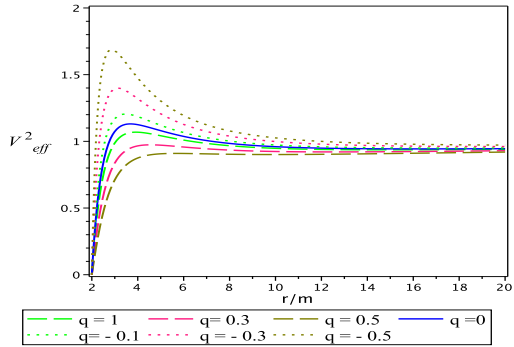}%
\caption{The effective potential for timelike  geodesics on the equatorial plane as a function of the radius for different values of the quadrupole parameter. Here we set $l^2=20$  for concreteness. }%
\label{fig2}%
\end{figure}

Figure \ref{fig2} illustrates the behavior of the effective potential in terms of the parameter $q$ for $\theta = \pi/2$.   
The effective potential of the Schwarzschild spacetime is also shown for comparison. For positive (negative) values of $q$, the effective potential at a given point outside the outer singularity is always less (greater) than the Schwarzschild value.  This indicates that the distribution of  orbits on the equator of the metric (\ref{apin1})  can depend  drastically on the value of $q$. 

\subsection{Circular orbits}
\label{sec:cir}

We will now investigate the properties of circular orbits on the equatorial plane, $\theta=\pi/2$ of the gravitational described by the approximate metric (\ref{apin1}). 
Circular orbits correspond to the limiting case $\dot r=0$. Their stability properties are determined by the extrema of the effective potential. 
Following the conventional stability analysis of circular orbits involving a potential function, from Eq.(\ref{veff}), we obtain 
\bea \nonumber
&V_{eff,r}=& 
\frac{2 \epsilon m}{r^2} + \frac{2 \left(3m-r \right)l^2 }{r^4} 
+\frac{qm}{r^2 \left(r-m \right)^2} \nonumber \\ 
&\times& \bigg[ 
\left(2 m^2 -4mr +r^2 \right) \epsilon \nonumber \\ &+& \frac{2\left(6m^2-10mr+3r^2 \right) l^2}{r^2}
\bigg] \ ,
\eea 
and
\be 
V_{eff,rr}= B_1 +B_{2}
\ee
where $B_{1}$ and $B_{2}$ are given by 
\be 
B_{1}= \frac{m \left( 36m^4 -90m^3 r+74m^2r^2-22mr^3+2 r^4\right)}{r^3 \left(3m^2-4mr +r^2 \right)^2} \ ,
\ee

\be 
B_{2}= \frac{mq \left( 36m^4 -114m^3 r+74m^2r^2-17mr^3+r^4\right)}{r^3 \left(3m^2-4mr +r^2 \right)^2}
\ee   
where we set $\epsilon=1$ and replaced the value of the angular momentum 
 \bea \label{l2co} \nonumber
&l^2&= \frac{\epsilon m r^2}{\left(r-m\right)\left(r-3m\right)^2} \bigg[ 
2 \left(3m^2-4mr+r^2\right) \\&+&q \left( r^2-6m^2\right) \bigg],
 \label{cir_l2}
 \eea
which can be derived from the condition $V_{eff,r} =0$.

\begin{figure}%
\includegraphics[width=7cm, height=7cm]{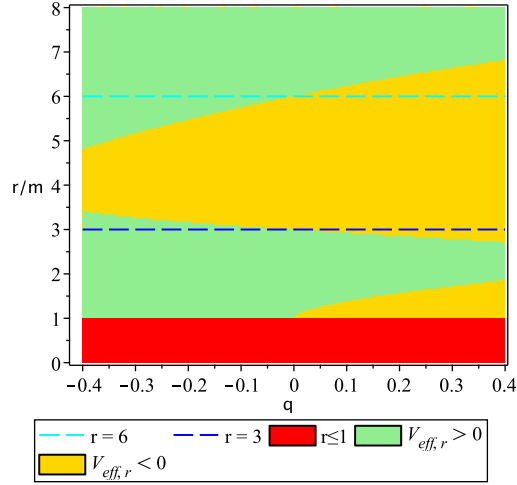}%
\caption{
Stability analysis of circular orbits with radius $r/m$ for different values of the quadrupole parameter $q$. Green (yellow) points represent stable (unstable) orbits. The red region lies inside the singularity. }%
\label{fig_sta_Vrr}%
\end{figure}

The numerical analysis of the stability condition, $V_{eff,rr} > 0$, is depicted in Fig.\ref{fig_sta_Vrr}. 
The green region contains only stable orbits whereas the yellow region corresponds to unstable orbits. For comparison, we include the limiting values of the Schwarzschild spacetime. 
We see that the quadrupole parameter $q$ changes the value of the minimum allowed radius ($3m$) and  of the inner most stable circular orbit radius ($6m$) of the Schwarzschild metric. In fact, the quadrupole leads to the appearance of a second stable region below and over the radius $3m$, which is not present in the Schwarzschild limiting case. This region can reach the value of $r/m \approx 1 $, approaching the singularity which is located at $r=m$. Moreover, within the spacetime determined by the interval $r\in (m, 2m]$, we notice that most of this region allows the existence of stable circular orbits with a disjoint region of instability for positive values of $q$.

\begin{figure}
\includegraphics[width=6cm,height=5cm,scale=0.1]{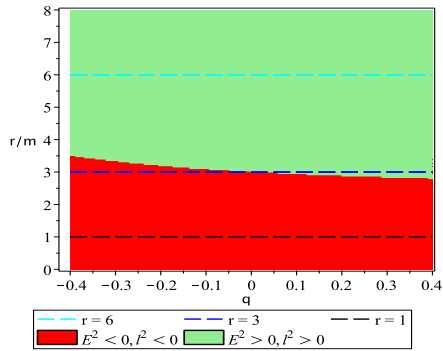}
\caption{ 
{Energy $E^2$ and angular momentum $l^2$ of test particles on circular orbits in terms of the radius $r/m$ and the quadrupole parameter $q$.
}
 }
 \label{fig_sta_E2l2}
\end{figure}

Furthermore, the energy of test particles on circular orbits can be expressed as
\be
 {E} ^2= \frac{\epsilon \left(r-2m \right)^2}{r\left( r-3m\right)} \bigg[
 1- \frac{mq \left(r-6m \right)}{2 \left( r-m\right)\left(r-3m \right)}
 \bigg].
\ee
In Fig. \ref{fig_sta_E2l2}, we plot the regions in which the energy $E^2$ and angular momentum $l^2$ are both positive or negative simultaneously. 
The red region denotes all the radii that are not allowed for circular orbits because either the squared of the energy or of the angular momentum are negative. 
A comparison with Fig. \ref{fig_sta_Vrr}
shows that the region contained  between the singularity $r/m=1$ and  around $r/m=3$ is allowed for circular orbits by the stability condition but is excluded by the energy and angular momentum conditions.

We conclude that the effect of the quadrupole on the properties of circular orbits is as follows. A positive quadrupole leads to an increase of the minimum allowed radius whereas a negative quadrupole generates the opposite effect. This means that only in the case of an oblate object, test particles are allowed to exist on orbits closer to the singularity, which is situated at $r=m$.

\begin{figure}
\includegraphics[width=6cm,height=5cm,scale=0.1]{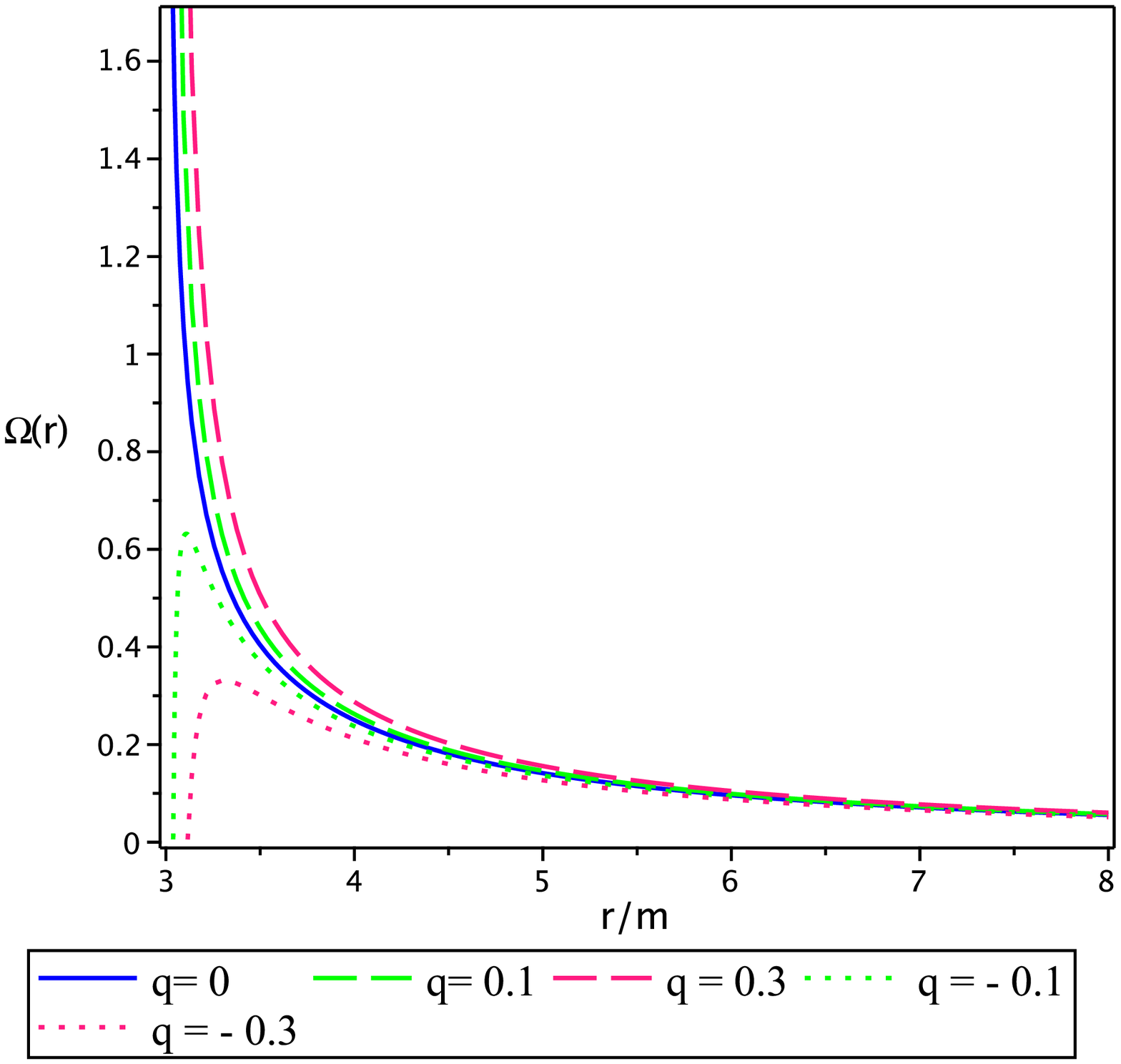}
\includegraphics[width=6cm,height=5cm,scale=0.1]{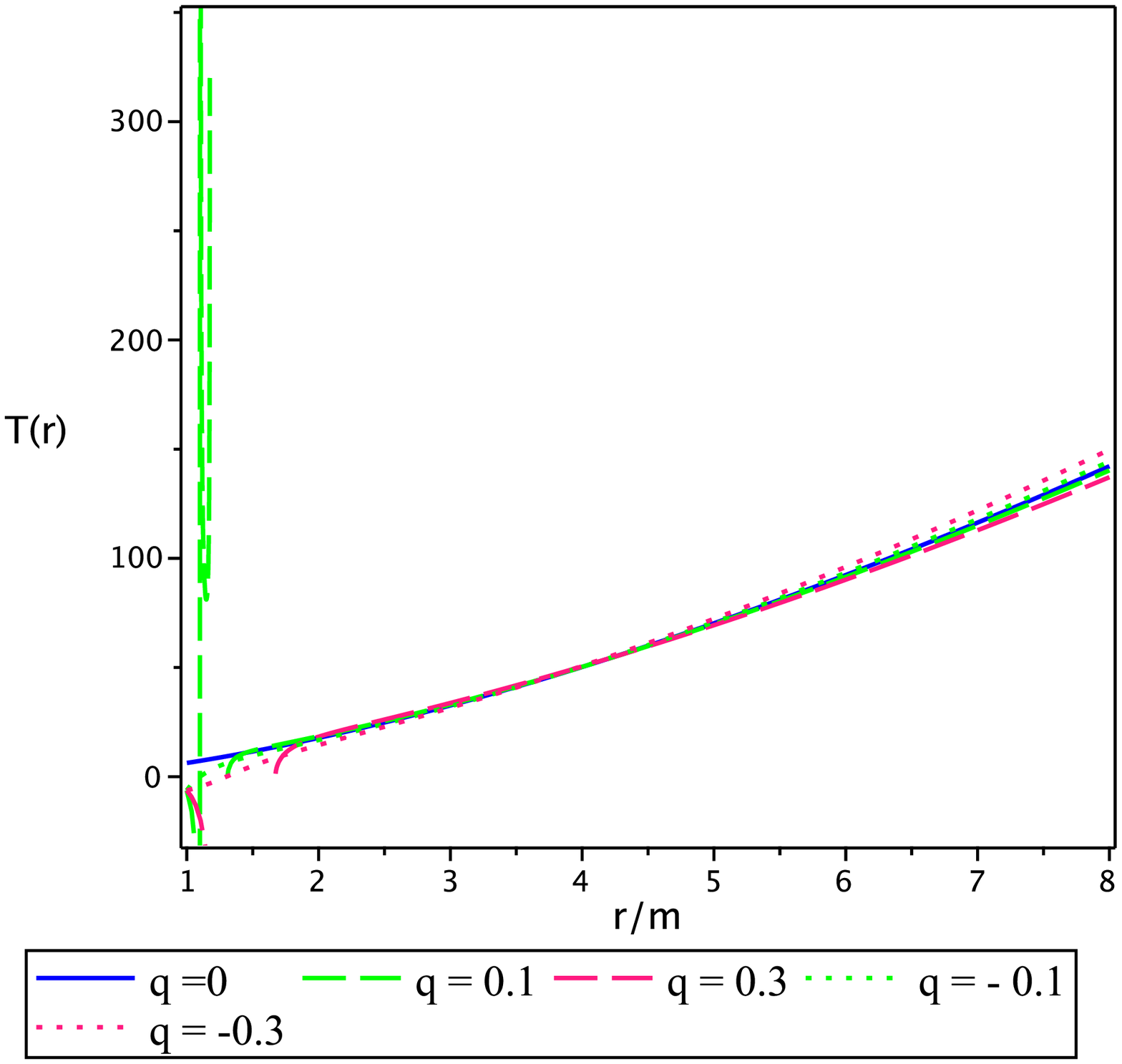}
\caption{Angular velocity and  period of circular orbits in terms of the radius $r/m$ and the quadrupole parameter $q$. }
\label{fig ometa_T}  
\end{figure}

Finally, we consider the  angular velocity
\bea \nonumber  
&\Omega(r)=& \dot\varphi= \frac{1}{r} \bigg[  \frac{m}{r-3m} \bigg]^{1/2}  \\ &&\times \bigg[ 1 + \frac{q \left( 6m^2-4mr+r^2\right)}{4 \left(r-m \right)\left(r-3m \right)} \bigg],
\eea 
and the period 
\bea \nonumber
T(r) &=& \int \frac{\dot t}{\dot\varphi} d\varphi = 2\pi \frac{dt}{d\varphi} \\ &=& 2\pi r^{3/2} m^{-\frac{1}{2}} \bigg[ 1- \frac{q \left(r-4m \right)}{4 \left(r-m \right)}\bigg],
\eea 
of circular orbits.  
%In  Fig. \ref{fig_sta_E2}, we illustrate the behavior of the energy and the angular momentum in terms of the radius and the quadrupole parameter.
%\begin{figure}
%\includegraphics[width=6cm,height=5cm,scale=0.1]{fig_sta_E2.eps}
%\includegraphics[width=6cm,height=5cm,scale=0.1]{fig_sta_l2.eps}
%\caption{Energy $E^2$ and angular momentum $l^2$ of test particles on circular orbits in terms of the radius $r/m$ and the quadrupole parameter $q$.}  \label{fig_sta_E2} \end{figure}
The behavior of the angular velocity and period are depicted in Fig. (\ref{fig ometa_T}). We can see that the influence of the quadrupole on the value of the angular velocity increases as the radius of the orbit approaches the value of $r\approx 3m$.  This agrees with the behavior of the stability condition and the energy and angular momentum shown in Figs. \ref{fig_sta_Vrr} and \ref{fig_sta_E2l2}, respectively. For a given orbit radius, the angular velocity increases (decreases) for positive (negative) values of the quadrupole. Notice that close to the singularity located at $r=m$, there is a region in which the angular velocity is a well behaved function of $q$ and $r$. This region corresponds to the stability region that was also found in Fig. \ref{fig_sta_Vrr}.

%%%%%%%%%%%%%%%%%%%%%%%%%%%%%%%
%%%%%%%%%%%%%%%%%%%%%%%%%%%%%%%

\subsection{Bounded and unbounded orbits}
\label{sec:unb}

\begin{figure}%
\includegraphics[scale=0.19]{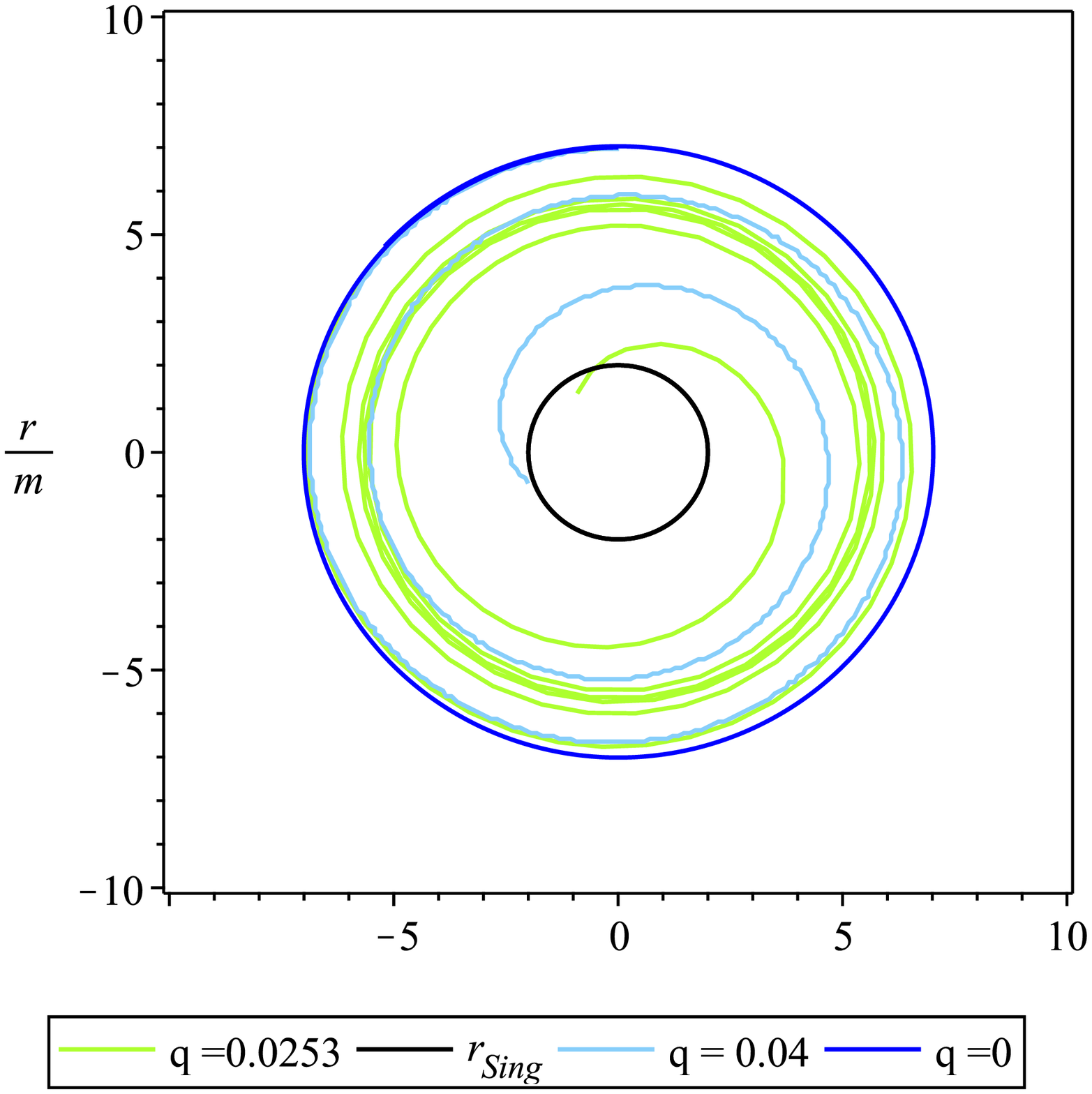}%
\quad 
\includegraphics[scale=0.19]{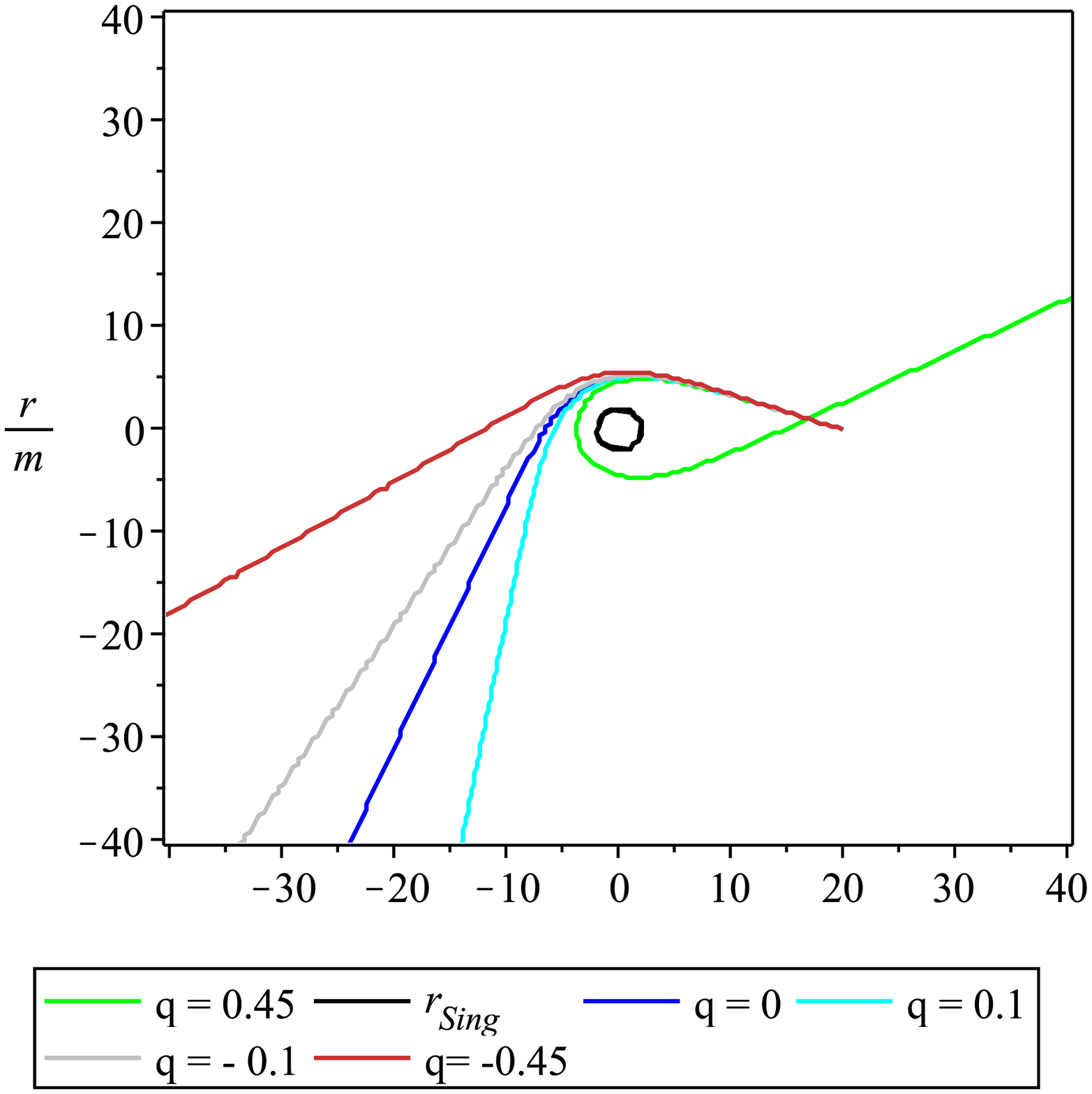}%
\quad 
\includegraphics[scale=0.19]{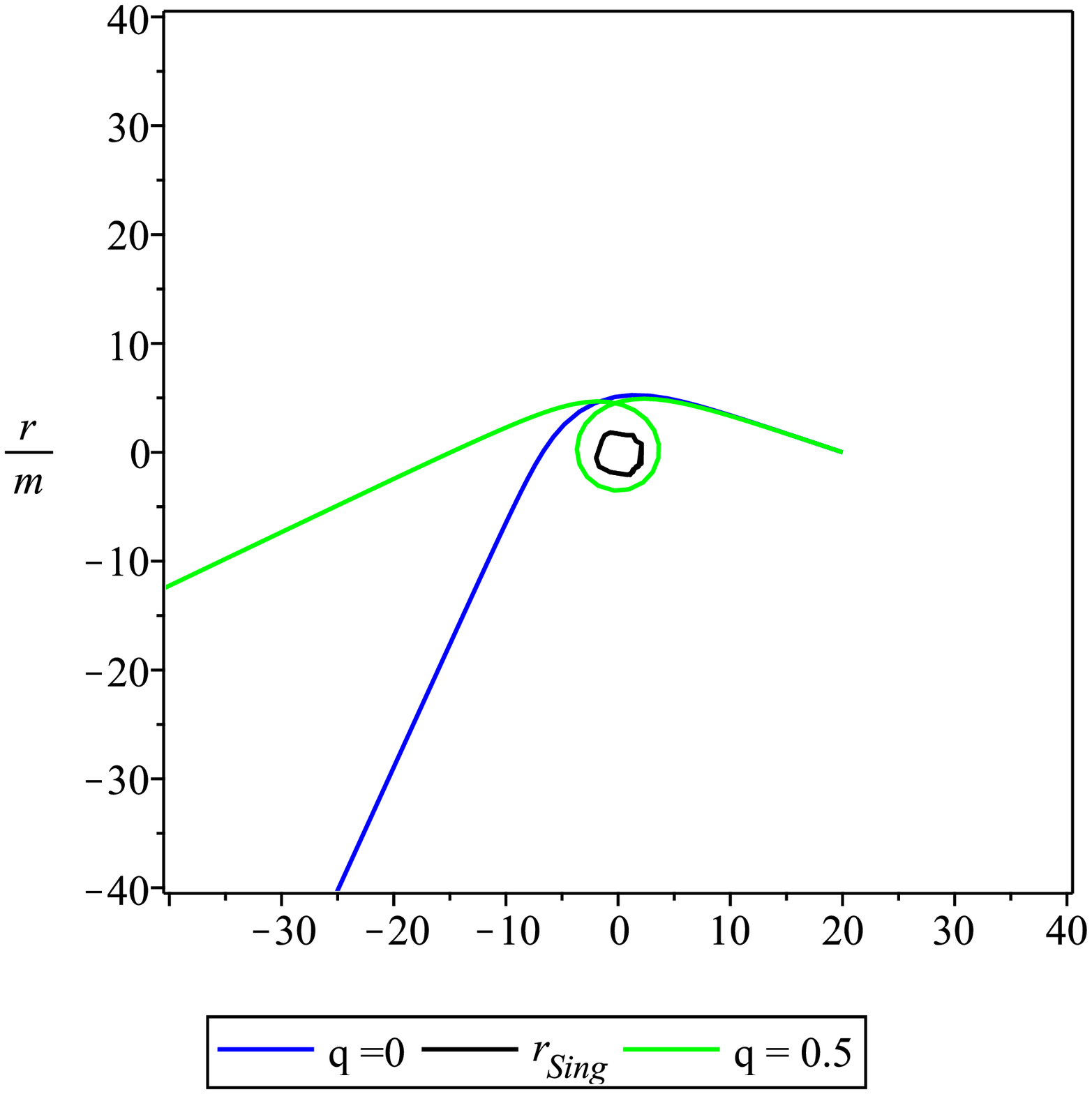}%
\caption{Influence of the quadrupole on unbounded orbits. 
Initial conditions: 
$ \varphi(0) = \pi/2,  \ r(0) =7, \ \dot{\varphi}(0)= 0.07145$, $\dot r(0)=0 $  (top, left);
$ \varphi(0) = 0,  \ r(0) =20, \ \dot{\varphi}(0)= 0.04305$, $\dot r(0)=-2.5 $ 
(top, right);
$ \varphi(0)=0 $, $r(0)=20$, $\dot{\varphi}(0)=0.04305$, $\dot{r}(0)=-2.489$ (bottom) ;} \label{figі3}
\end{figure}

\begin{figure}%
\includegraphics[scale=0.19]{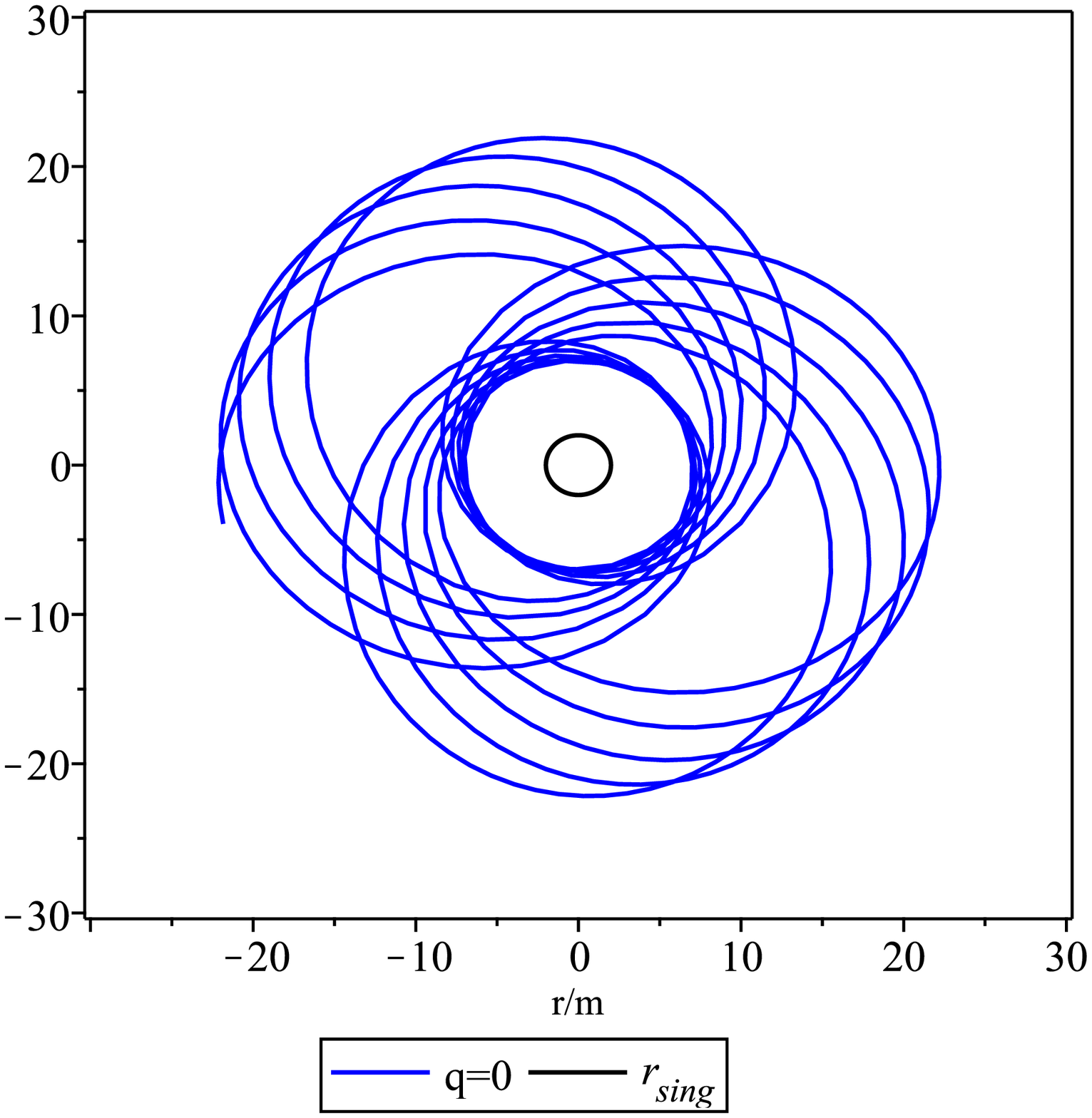}%
\quad
\includegraphics[scale=0.19]{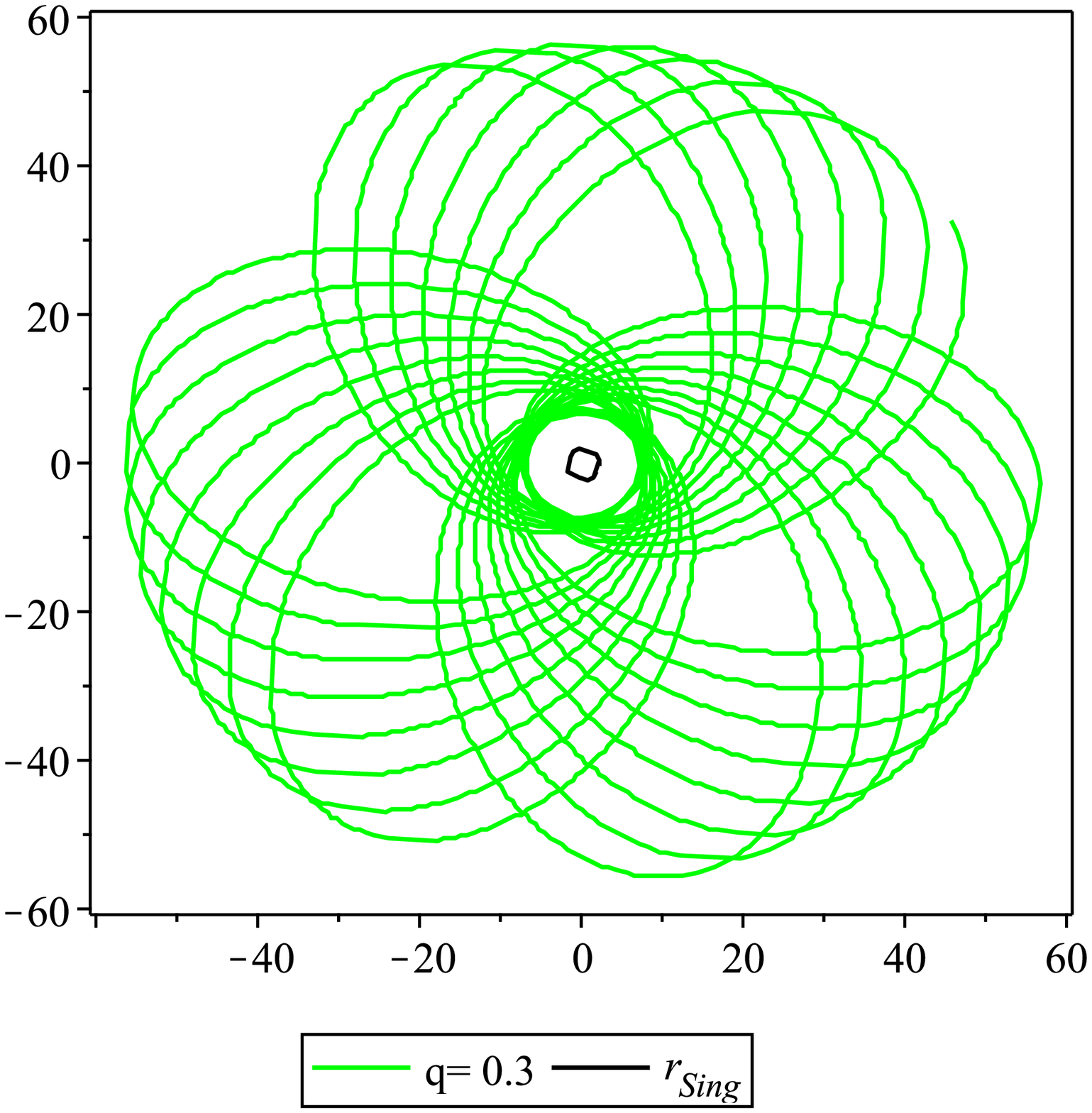}%
\quad
\includegraphics[scale=0.19]{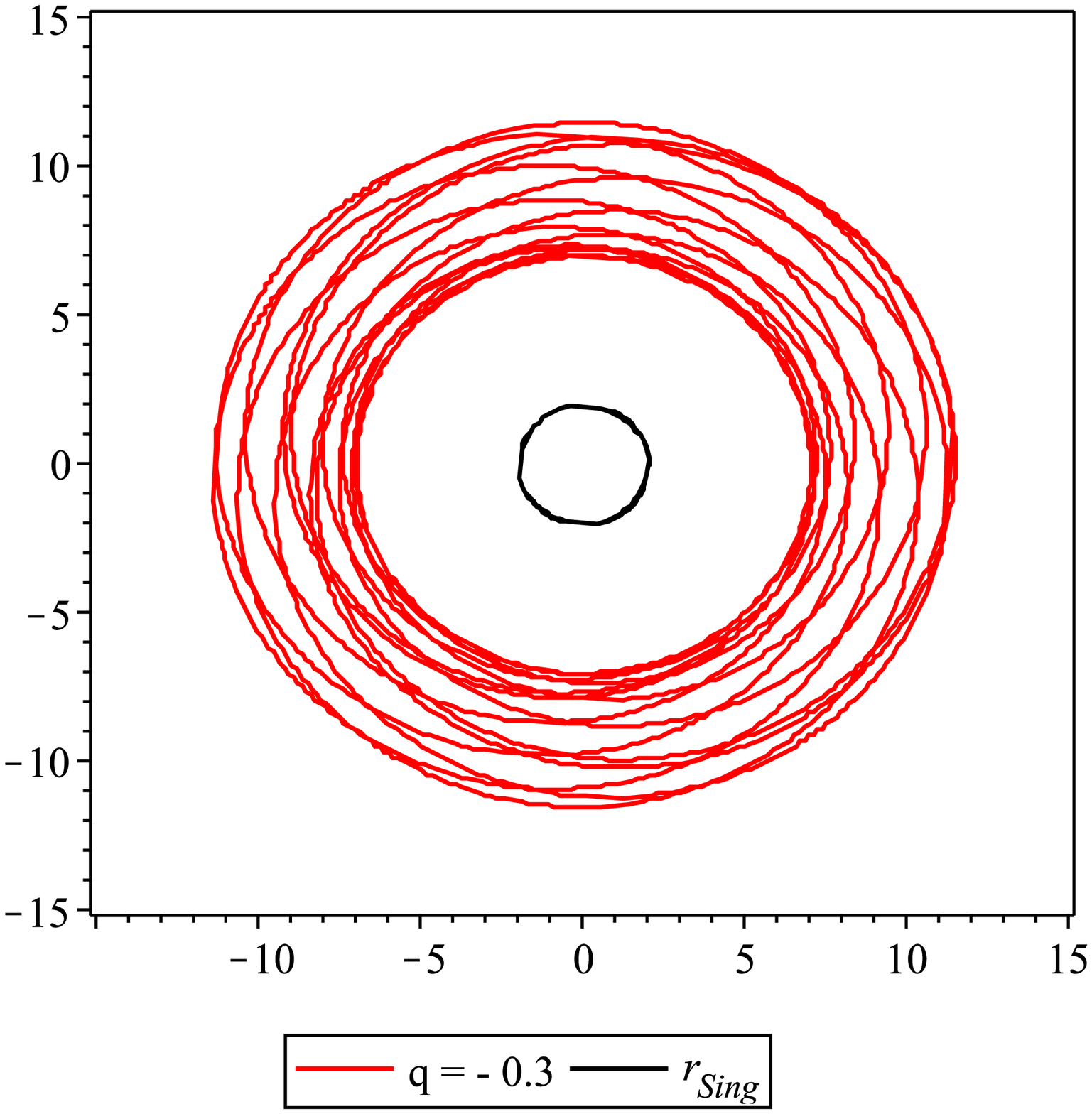}%
\caption{Influence of the quadrupole on Schwarzschild bounded orbits with vanishing initial radial velocity ($\dot r(0)=0)$.   
The initial conditions are  $\varphi(0)=0$, $r(0)=7$,     and $\dot{\varphi}(0)=0.08$ for all the trajectories. }
\label{fig4c}%
\end{figure}

We now study the influence of the quadrupole on the trajectories of massive test particles, moving along unbounded paths on the equatorial plane. 
  The geodesics for different values of the quadrupole parameter are given in Figs.  \ref{figі3}-\ref{fig4c},
  where the radial coordinate is dimensionless ($r/m$).
  
We consider first unbounded Schwarzschild trajectories  with non-zero initial radial velocities under the influence of the quadrupole.
In this case, we see that for the chosen initial angular and radial velocities all the  particles escape from the gravitational field of central slightly deformed body. This is illustrated Fig.\ref{figі3}.   
The direction along which the particle escapes to infinity depends on the value of the quadrupole. It is worth noticing that, in principle, this effect could be used to measure the quadrupole of the central mass distribution.

In Fig. \ref{fig4c}, we consider a Schwarzschild bounded orbit 
with zero initial radial velocity and the same non-zero
value for the initial angular velocity. 
The left panel shows the Schwarzschild geodesic. The  central and right panels illustrate the influence of a negative and positive small quadrupole, respectively.
We conclude that that the small quadrupole does not affect the bounded character of the geodesic, but it does drastically modify the morphology of the trajectories. 

We conclude that the quadrupole always affects the Schwarzschild trajectories. The explicit modifications depend on the properties of the original Schwarzschild trajectory and the value of the quadrupole parameter. 

\subsection{Radial geodesics and repulsive gravity}
\label{sec:rad}

We now study the free fall of test particles. To this end, we consider the geodesics equations in their most general form as given in Eqs.(\ref{eq_time})-(\ref{eq_phi}). The important point is that in all the cases to be considered all the initial spatial velocities are assumed to vanish. The starting point can be chosen arbitrarily, but some special values of the angle $\theta$ are of interest, namely, the symmetry axis $\theta=0$, the equatorial plane $\theta=\pi/2$, and some other different value that we can choose as $\theta=\pi/4$. In fact, in \cite{Boshkayev2016}, it was shown that by analyzing the behavior of radial geodesics one can detect the presence of repulsive gravity \cite{defelice89,lq12,lq14}

 A free falling particle will continue its motion along the radial direction, unless a force acts on it and changes the original radial direction. This phenomenon has been reported in the case of an exact solution with quadrupole moment in \cite{Boshkayev2016}.

\begin{figure}%
\includegraphics[width=3.7cm, height=4cm]{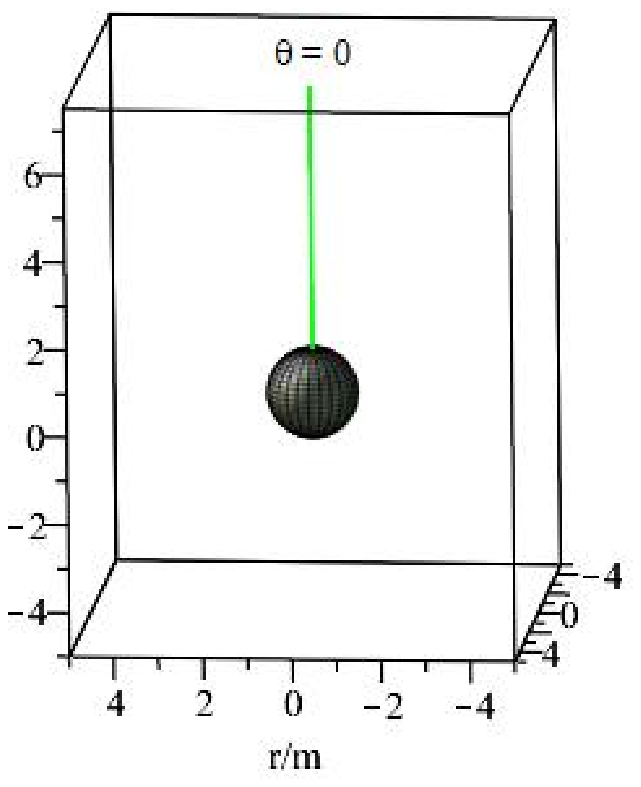}% 
\quad
\includegraphics[width=3.7cm, height=4cm]{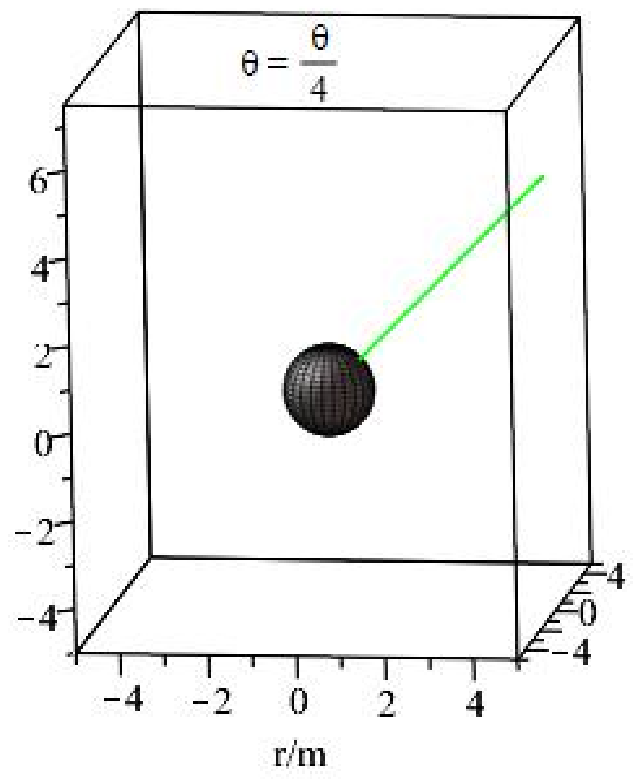}% 
\quad
\includegraphics[width=3.7cm, height=4cm]{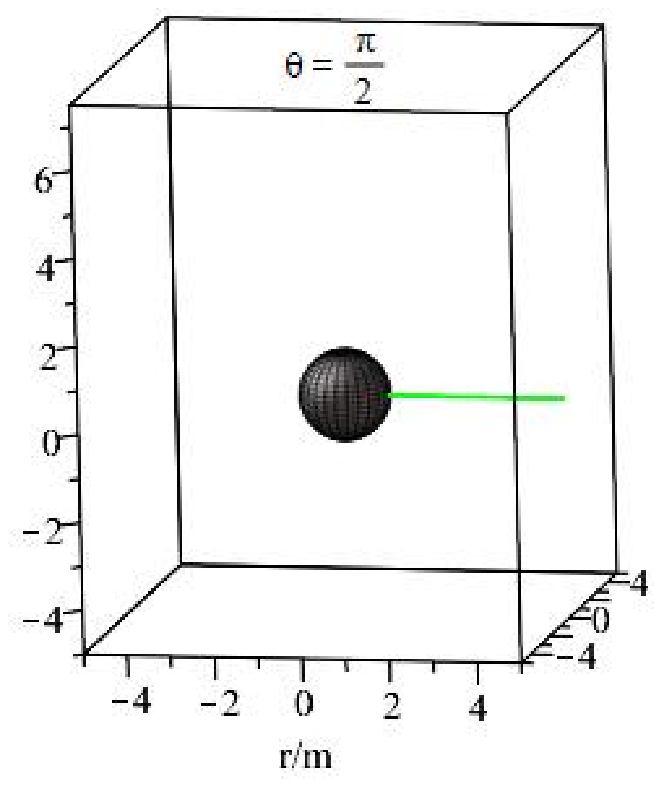}%
\caption{Free fall of test particles with vanishing initial velocities. On the axes $\theta = 0$ (top, left), $\theta = \pi/4$ (top, right), and $\theta = \pi/2$ (bottom). The sphere represents the singularity at the radius $r=m$ (top). 
The trajectories of the free falling test particles do not depend on the value of the parameter $q$.
 For concreteness, we set  the energy of the test particle as $E=1$. 
}
\label{fig_radial_3d}%
\end{figure}

We will now consider the same situation in the case of the approximate solution we are analyzing in this work. The result of the integration of the geodesic equations 
(\ref{eq_time})-(\ref{eq_phi}) is illustrated in Fig. \ref{fig_radial_3d}.  We see that, in fact, free falling particles move along the original directions ($\theta =0$, $\theta=\pi/4$, and $\theta=\pi/2$), independently of the value of quadrupole parameter $q$. This means that no repulsive gravity effects can be detected in the case of the approximate metric. Taking into account also the result of \cite{Boshkayev2016}, we conclude that repulsive effects are non-linear, i.e., they appear only in the case of an exact quadrupolar metric.

%\begin{figure}%
%\includegraphics[scale=0.19]{radial_fig12_prd.eps}%
%\quad
%\includegraphics[scale=0.19]{radial_approxi_qmet.eps}%
%\quad
%\caption{fig.12-PRD(left) and for approximate q metric(right). }
%\label{radial_fig12_prd}%
%\end{figure}

%%%%%%%%%%%%%%%%%%%%%%%%%%%%%%%%%%%%%%%%

%%%%%%%%%%%%%%%%%%%%%%%%%%%%%%%%%%%%%%%%%%%%%%%%%%%%%%%%%%
%%%%%%%%%%%%%%%%%%%%%%%%%%%%%%%%%%%%%%%%%%%%%%%%%%%%%%%%%%%
\section{\label{sec:con}Conclusions and remarks}

In this work, we have derived a family of approximate vacuum solutions of Einstein equations with quadrupole. Among all the solutions contained in this family we choose a particular one, which presents a naked singularity at the hypersurface $r=m$, instead of $r=2m$ as in other quadrupolar metrics. To our knowledge this is the only metric with such a singularity.

By applying an appropriate coordinate transformation, we found the Newtonian limit of the approximate solution and showed that it corresponds to the gravitational potential of a mass with quadrupole.

Then, we investigated the motion of test particles along circular orbits. We established that a positive quadrupole leads to an increase of the minimum allowed radius whereas a negative quadrupole generates the opposite effect. This means that only in the case of an oblate object, test particles are allowed to exist on orbits closer to the singularity, which is situated at $r=m$. In the case of bounded and unbounded trajectories, we found that 
the quadrupole always affects the Schwarzschild trajectories. The explicit modifications depend on the properties of the original Schwarzschild trajectory and the value of the quadrupole parameter. 

Finally, we analyzed radial geodesics that have been used previously to detect effects of repulsive gravity in an exact quadrupolar metric. However, in the case of the approximate solution no repulsive effects were found. We conclude that repulsive gravity in the the quadrupolar naked singularities is a non-linear phenomenon.

\section*{Acknowledgments}

This work was partially supported  by Ministry of Education and Science (MES) of the Republic of Kazakhstan (RK), Grant No. BR10965191, and by UNAM-DGAPA-PAPIIT, Grant No. 114520, 
Conacyt-Mexico, Grant No. A1-S-31269. 
S.T. acknowledges the support through the postdoctoral fellowship program of Al-Farabi Kazakh National University.  
%\end{acknowledgments}

%\section{\label{app_linfield}Appendixes}

\appendix

%\nocite{*}
%\bibliographystyle{spbasic}

\bibliographystyle{spphys}
\bibliography{app.bib}
\end{document}